\documentclass[
  a4paper,
  fontsize=11pt,
  captions=tableheading,
  parskip=never,
  ]{scrartcl}
\setkomafont{caption}{\small}
\setkomafont{captionlabel}{\bfseries}
\addtokomafont{publishers}{\large}
\addtokomafont{subject}{\mdseries\large}

\pdfoutput=1

\usepackage{ILD}
\usepackage{graphicx}

\usepackage[utf8]{inputenx}
\usepackage[T1]{fontenc}
\usepackage[british]{babel}
\usepackage{csquotes}

\usepackage{subcaption}
\usepackage{import}
\usepackage{xspace}
\usepackage{graphicx}
\usepackage[shortcuts]{extdash}
\usepackage{siunitx}
\DeclareSIUnit \lightspeed {\text{{c}}}
\usepackage{xcolor}
\definecolor{linkblue}{HTML}{264772}
\usepackage{hyperref}
\hypersetup{
    colorlinks=true,
    linktocpage=true,
    linkcolor=linkblue,
    citecolor=linkblue,
    urlcolor=linkblue
}

\usepackage{wrapfig}

\graphicspath{ {./Pictures/} }
\DeclareGraphicsExtensions{.pdf,.png,.jpg}

\begin{document}

\hyphenation{
  am-pli-fi-ca-tion
  col-lab-o-ra-tion
  per-for-mance
  sat-u-rat-ed
  se-lect-ed
  spec-i-fied
}

\title{CPID: A Comprehensive Particle Identification Framework for Future e$^+$e$^-$ Colliders}
\author{Ulrich Einhaus}

\titlecomment{Talk presented at the International Workshop on Future Linear Colliders (LCWS 2023), 15-19 May 2023. C23-05-15.3.}
\date{}

\addauthor{Ulrich Einhaus}{\institute{1}}
\addinstitute{1}{Deutsches Elektronen\-/Synchrotron DESY, Germany}

\abstract{
With the broadening landscape of proposals for future Higgs, top and electroweak physics factories, detector diversity as well as the reach and depth of physics analysis increase.
One emerging topic of renewed interest is particle identification (PID).
This paper highlights the available technology options and the physics need for dedicated PID.
It introduces a new framework to perform a coherent PID assessment across the different future collider proposals, called Comprehensive PID (CPID).
Its structure is laid out, and examples are shown, which demonstrate the power and flexibility of this approach.
}

\titlepage
\clearpage

\tableofcontents
\newpage

\section{Introduction}
While the international particle physics community agrees that the next big collider should be an e+e- Higgs factory, it is very much open which project will be realised.
Several accelerator concepts have been proposed, with the most notable differences between linear and circular ones, and even more collider detectors are "on the market", with various stages of simulation detail.
Since the resources for these future collider projects are still very limited, there is a significant interest to work on common tools, which can be used for any and all of them.
This concerns in particular software tools, which are generally easily transferable, and the decision was taken to use a common framework, called key4HEP \cite{key4HEP}.
In order not to start all over, interfaces were created to utilise existing and successful software as much as possible.
One example of this is the inclusion of iLCSoft \cite{lcsoft,ilcsoft_git}, which has been successfully used by the linear collider community for many years.
The central simulation and reconstruction tools of iLCSoft have been made available in key4HEP via a wrapper.

\begin{figure}[!htp]
  \centering
  \begin{subfigure}{0.48\textwidth}
    \includegraphics[width=\textwidth,keepaspectratio=true]{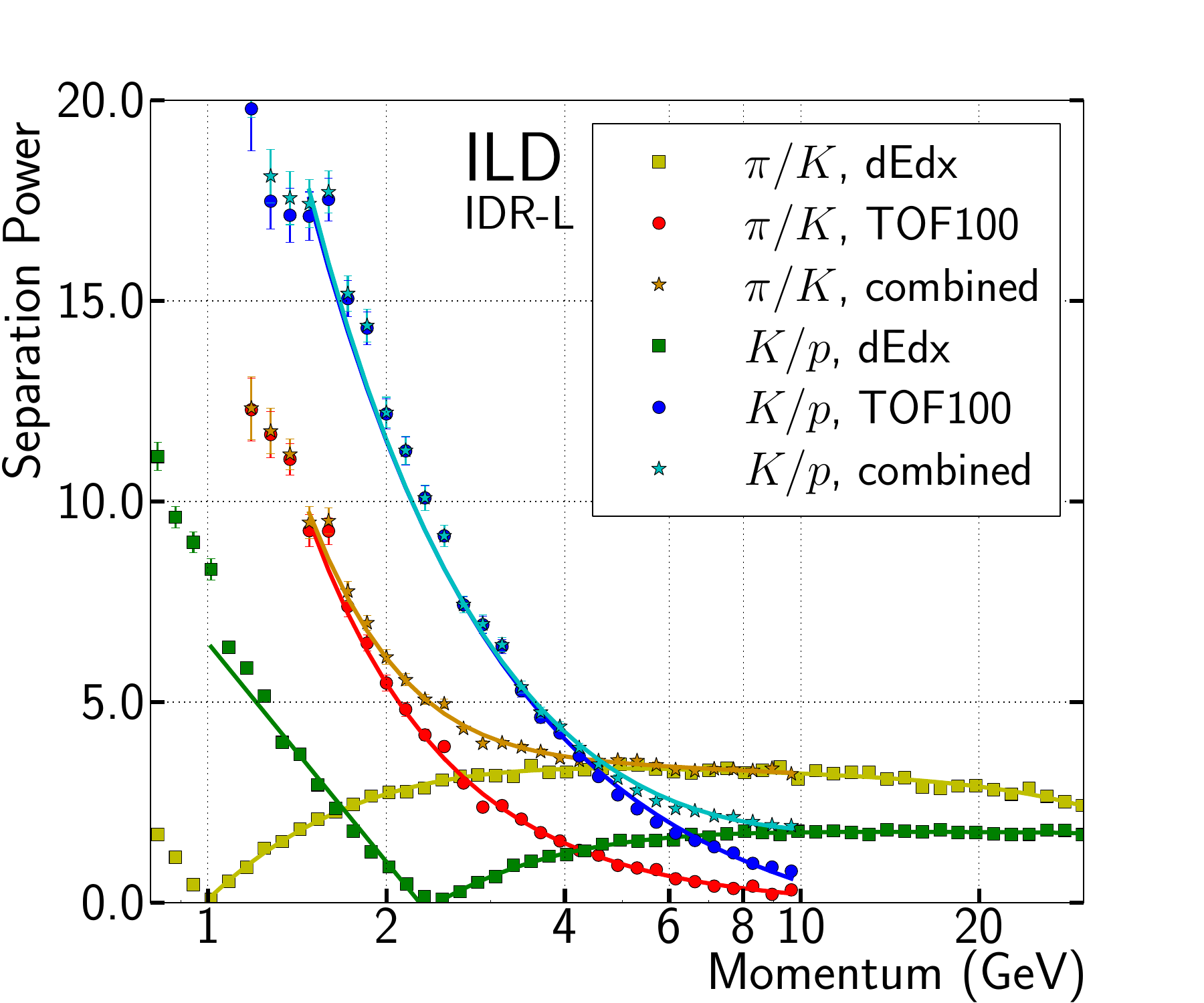}
    \caption{ILD TPC dE/dx and TOF separation power, from \cite{ILD_IDR}.}
  \end{subfigure}
	\hspace{.02\textwidth}
  \begin{subfigure}{0.48\textwidth}
    \vspace{2cm}
    \includegraphics[width=\textwidth,keepaspectratio=true]{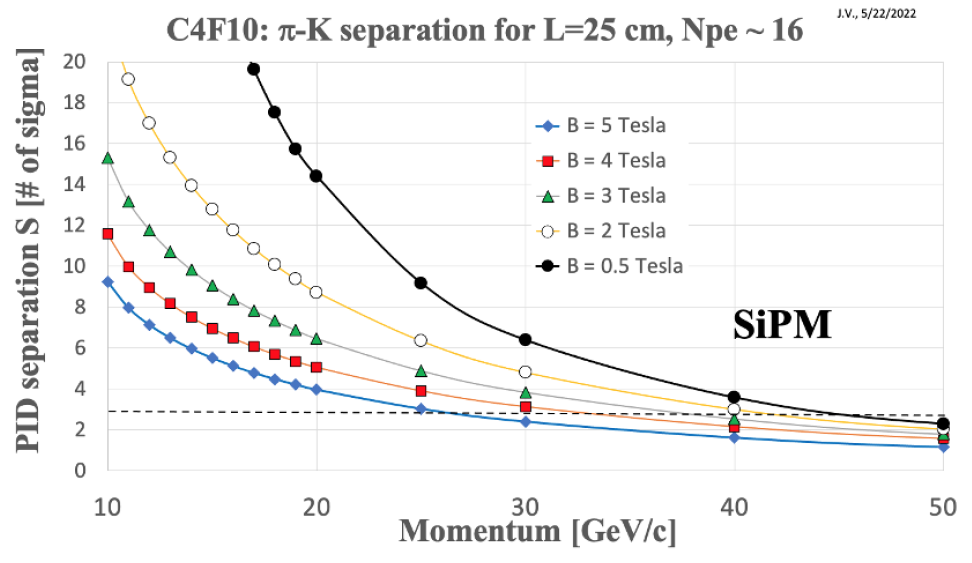}
    \caption{RICH separation power, from \cite{Hssbar}.}
  \end{subfigure}
  \caption{Typical pion-kaon separation power curves based on simulation with different underlying technologies.}
  \label{fig:PID_Perf_Exm}
\end{figure}

A common more physics-related topic of the future collider proposals is particle identification (PID).
This refers generally to the identification of the species of detector-stable particles, but often more specifically to hadron ID as well as low-momentum electron and muon ID via dedicated PID-capable subdetectors.
The interest in PID has increased in recent years, both because technological advances allow better PID performance, as well as because its impact on physics has become clearer with ongoing studies.
In particular the advent of circular collider proposals, which necessarily concentrate on lower center-of-mass (com) energies compared to linear colliders, has sparked interest:
On the one hand, at lower com energies also the momenta of individual particles in the detector are lower on average, which makes PID systems whose performance typically degrades with increasing $\beta\gamma$ more effective, compare \autoref{fig:PID_Perf_Exm}.
On the other hand, a Tera-Z programme allows to study B physics in channels which are unavailable at the current B factories, for which an excellent PID is plainly necessary.

This study addresses the need for a common software tool called "Comprehensive Particle Identification" (CPID) for PID at future Higgs factories.
It introduces a modular framework that allows to assess, combine and apply PID to the different collider concepts in a coherent way.
In chapter 2, the underlying motivation for this study is elaborated on, summarising examples of physics processes which profit from or depend on PID information as well as the different PID technologies being discussed.
Chapter 3 introduces the concept and structure of the CPID tool, before exemplary results of assessment and combination of different technologies are highlighted in chapter 4 and chapter 5 concludes.

\section{From PID Technology to Analysis}

The PID systems currently under discussion for future Higgs/top/electroweak (HTE) factories are based on three different effects that depend on $\beta\gamma$: ionisation, Cherenkov radiation and speed of propagation.

\begin{figure}[!htp]
  \centering
	\begin{subfigure}{0.5\textwidth}
    \includegraphics[width=\textwidth,keepaspectratio=true]{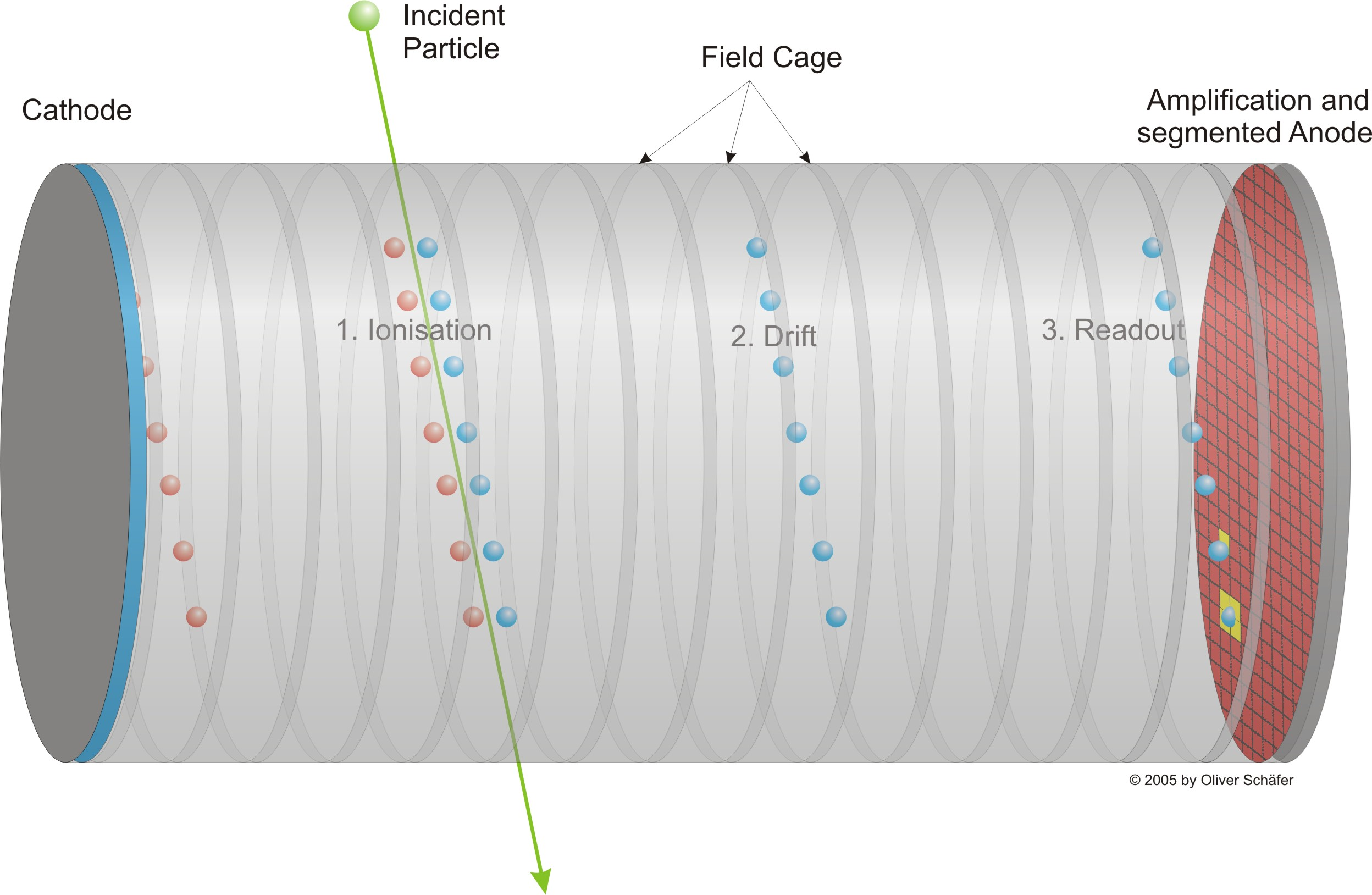}
    \caption{TPC scheme, credit: O. Sch\"afer.}
		\label{fig:TPC_TOF_left}
  \end{subfigure}
  \hspace{.1\textwidth}
  \begin{subfigure}{0.35\textwidth}
    \includegraphics[width=\textwidth,keepaspectratio=true]{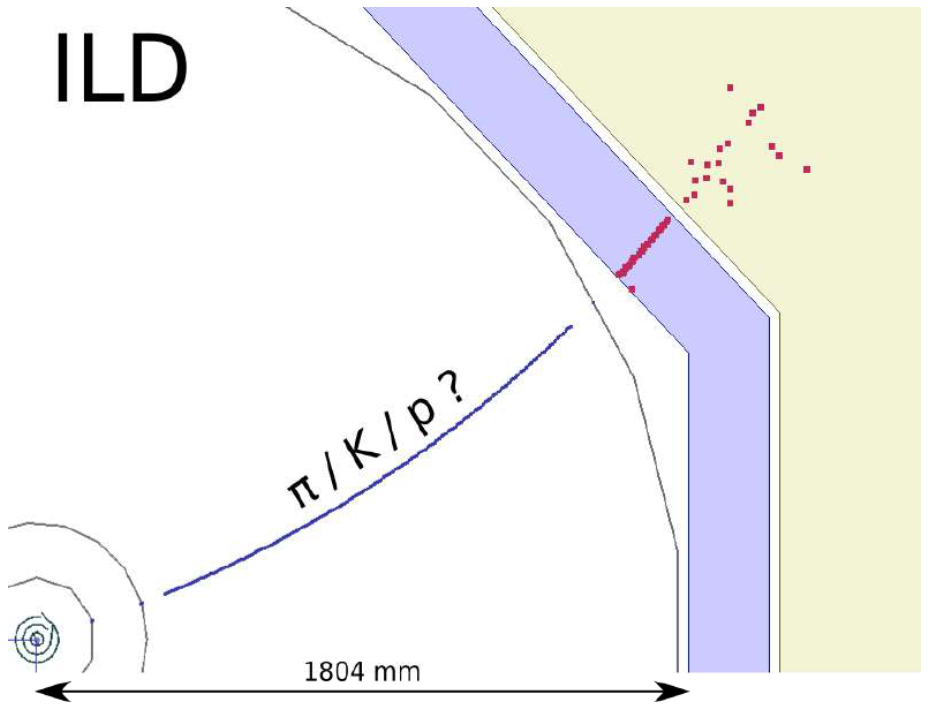}
    \vspace{.2cm}
    \caption{TOF scheme, from \cite{TOF_Talk}.}
		\label{fig:TPC_TOF_right}
  \end{subfigure}
  \caption{Usage of the gaseous tracker for dE/dx and the ECal for TOF.}
  \label{fig:TPC_TOF}
\end{figure}

The specific energy loss of a particle from ionisation, or dE/dx, is following the Bethe-Bloch curve, which can be used for identification. Each ionising interaction is governed by a Landau distribution with a long tail to high energies. Therefore, to calculate an average energy loss, usually a truncated mean is used by truncating a fixed fraction of all individual measurements that contribute to the dE/dx of a track. This becomes more effective with a larger number of measurements, which is why dE/dx has been used for decades in gaseous detectors with O(100) measurements per track (\autoref{fig:TPC_TOF_left}), while the corresponding resolution from dE/dx in O(10) layers of solid state detectors (in particular Si trackers) does not suffice for a good PID. Recent developments in gas ionisation PID concentrate on increasing granularity in space or time in order to resolve the number of ionising interactions, which follows a curve similar to Bethe-Bloch but is governed by Poisson statistics and thus has typically about a factor 2 better effective resolution.

Also Ring Imaging Cherenkov (RICH) detectors have been used for many years.
Here, a particle creates Cherenkov radiation in a traversed material, which is then imaged to a readout plane, with the reconstructed Cherenkov angle being correlated to $\beta\gamma$.
The larger the refractive index, the lower the minimum momentum necessary to create Cherenkov radiation, but the lower also the maximum momentum at which effective separation is possible. Therefore, for different ranges of expected momenta of particles in a collider, different radiation materials are used: While Belle 2 uses a quartz radiator for particles up to 5 GeV, RICH systems proposed for a Higgs factory utilise gas and/or aerogel radiators.
In the past, the imaging systems typically consisted of large parabolic mirrors and a photomultiplier tube array. This takes a lot of space and was/is used mostly in collider detectors which cover only a small solid angle with respect to the interaction in the lab frame (e.g.\ HERA-B or LHCb). There, the readout was/is placed outside the sensitive solid angle.
Silicon photomultipliers, however, allow for a compact and reasonably low-material readout plane inside the active volume, enabling RICH systems in both endcap and barrel region of a $4\pi$-detector.

\begin{figure}[!htp]
  \centering
  \begin{subfigure}{0.45\textwidth}
    \includegraphics[width=\textwidth,keepaspectratio=true]{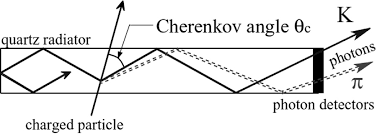}
    \caption{TOP scheme, from \cite{TOP_Paper}.}
  \end{subfigure}
  \hspace{.05\textwidth}
  \begin{subfigure}{0.45\textwidth}
    \includegraphics[width=\textwidth,keepaspectratio=true]{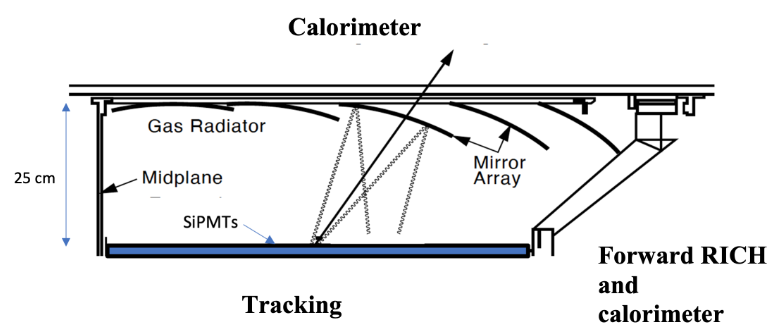}
		\caption{RICH scheme, from \cite{Hssbar}.}
  \end{subfigure}
  \caption{Cherenkov-based PID systems.}
  \label{fig:TOP_RICH}
\end{figure}

Finally, the arguably simplest way of determining the mass of a particle and thus identifying it, is measuring its absolute velocity and combining it with its momentum.
This can be done using time-of-flight (TOF) systems, which measure the arrival time of a particle, see \autoref{fig:TPC_TOF_right}. The necessary timing resolution of {< 100 ps} has become available in recent years as they were developed for pile-up rejection in ATLAS and CMS.
TOF systems are usually placed between the (outer) tracker and the calorimeter, since the longest possible lever arm is desired before most particles undergo hard scattering in the calorimeter.
To determine the absolute velocity, the flight time and the track length are needed. The time in e$^+$e$^-$ colliders is generally measured relative to the bunch clock, since the interaction areas are small compared to the achievable time resolution. The track length needs to be precisely reconstructed, taking ionisation and scattering in the inner detector systems into account. Depending on the tracking system and the envisioned timing precision, both aspects can be limiting factors to the velocity measurement.
\\

Most dedicated applications of PID use the distinction between pions and kaons, i.e.\ between down and strange quark, for flavour-related physics.

The first example is the reconstruction of exclusive channels in B physics. This becomes relevant for an HTE factory at the Z pole, where - in contrast to SuperKEKB/ Belle II - the com energy is sufficient to generate $B_s$ mesons, and the luminosity is sufficient together with very clean events to exceed the analysing power of LHCb.
\cite{Bs_Talk} shows that in $B_s \rightarrow D_s K$ PID has a massive impact on the suppression of the much more prevalent $D_s \pi$ background. Here, the PID up to about 35 GeV is required and needs to be excellent, since for a successful process identification every single hadron must be correctly identified, in processes that often have 3 or more hadrons.

Another large area of PID application is flavour tagging. Here one uses the fact that s quarks generate more strange hadrons and with higher momentum than d or u quarks during hadronisation.
Similarly, the b and c quarks of B and D mesons typically decay to s quarks, whose associated strange hadrons can be identified, in addition to the secondary or tertiary vertices.
In the end, the entire spectrum of kaons, but also pions and protons, can then be correlated to the initial quark flavour from the hard interaction.
Two analyses that use this are \cite{Malek21} and \cite{Einhaus21}, which study hadronic Z and W decays, respectively, in order to measure the couplings of the electroweak bosons to the different quark flavours.
\cite{AFB_Overview} shows an overview of $A_{FB}$ studies, i.e.\ the measurement of the forward-backward asymmetry of $q\bar{q}$ pairs in the s-channel.
Since the quark flavours have different asymmetries, each flavour is studied exclusively, utilising a strong double-tagging method since here both quark and antiquark have the same flavour. The direction of the initial (anti) quark is then determined by either a vertex charge measurement or by correlating it with outgoing (anti) kaons, which need to be correctly identified.
An explicit strange tagger has been implemented to study the potential of measuring the strange Yukawa coupling at a Higgs factory, see \cite{Hssbar}.

Moreover, TOF can be used directly to measure the mass of the charged kaon, an open question since the early 90s \cite{Kaon_Mass}.
Knowing the mass of a particle also allows to do a full track refit with correct energy loss, which improves the track parameters \cite{Track_Refit}.
This indicates that it may be worthwhile to re-assess the entire event in particle flow after PID.

A more general overview of various analyses and methods that already employ PID can be found in \cite{PID_Overview}.

\section{CPID Concept and Structure}

The different technological options and the broad application spectrum lead to a non-trivial optimisation problem regarding the inclusion in future colliders. Some central questions are:
How can we best combine PID technologies to achieve the best physics results? How can we assess the PID performance on an intermediate abstract level in a way that applies to all technologies and is still meaningful for analyses? What methods can be used for the best combination of PID information?

The following more detailed example questions may serve to further guide the discussion:
At what timing resolution does TOF start to be relevant for flavour tagging?
How does \textbf{my} physics channel depend on the dE/dx (dN/dx) resolution?
What if we use a silicon tracker and a RICH instead of a gaseous tracker in a given detector concept?
\\

The proposed solution for these questions is a new software framework called 'Comprehensive Particle Identification', or CPID.
Its core concept is modularity, both in the PID assessment algorithms as well as in the models that combine the PID information.
These modules can then be easily switched, while the core code basically takes care of book keeping. In addition, simple data structures for storage and interfaces mean low barriers for developers.

The CPID framework is for now being implemented in LCIO \cite{lcio} and Marlin \cite{marlin} within iLCSoft, which is part of key4HEP. Via an existing wrapper, it is immediately usable inside key4HEP-native EDM4HEP \cite{EDM4HEP} and Gaudi \cite{Gaudi}, and it should be transferred to these frameworks properly in the future.
The studies with CPID have so far been carried out with the ILD detector model.
In the standard reconstruction chain of ILD, CPID could serve as a replacement for the current PID assessment via the so-called LikelihoodPIDProcessor.
\\

\begin{figure}[!htp]
  \centering
  \includegraphics[width=\textwidth,keepaspectratio=true]{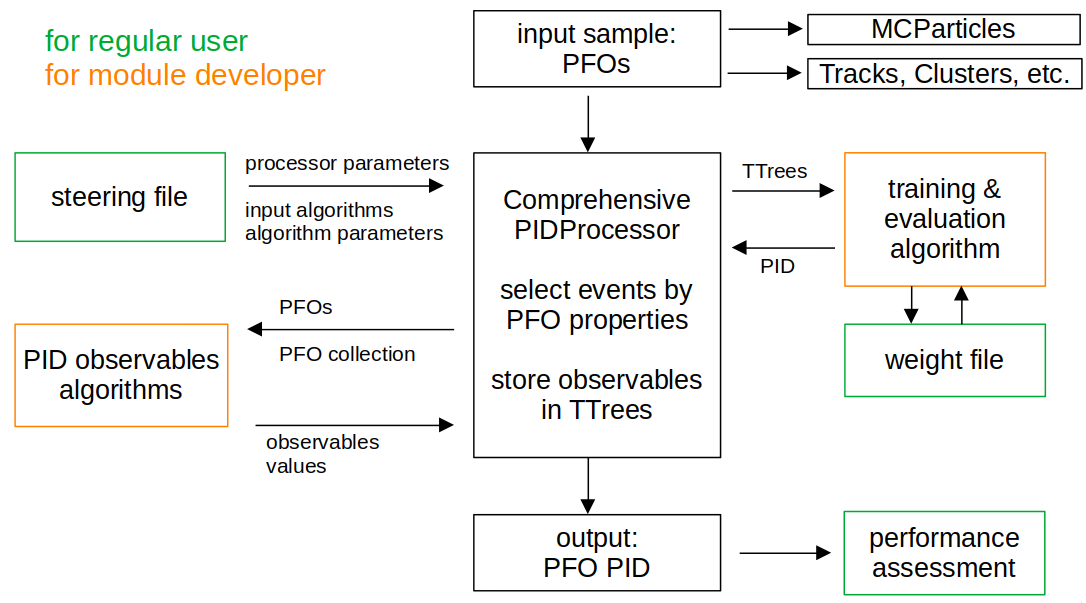}
  \caption{Structure of CPID.}
  \label{fig:CPID_Structure}
\end{figure}

An overview of the structure of CPID can be seen in \autoref{fig:CPID_Structure}.
The central element, called a processor in Marlin, takes particle flow objects (PFOs), which gives indirect access to the PFO constituents, i.e.\ tracks and calorimeter clusters, as well as the related Monte-Carlo particles.
The processor is handled via a steering file which gives access to its parameters as well as to the ones of the modules.
It makes an event selection based on optionally specified cuts.
The PFOs are then transferred to the PID algorithm modules, which extract their corresponding PID observables. These are returned to the central processor and stored in a TTree. The ROOT TTree was chosen since it widely available and known, and many operations can be executed on it via predefined functions.
The PID assessment based on the observables is done via training and evaluation algorithms modules. The training is performed after the PID info all events of the training sample have been extracted. With TTrees, the ROOT TMVA package allows for easy implementation of multivariate analysis tools such as boosted decision trees (BDTs), but any tools can be used here. To use complex machine learning algorithms, the TTree can also be written to a root file in order to do the training outside of Marlin.
The result of training are corresponding weight files.
The inference is performed in a second iteration of the framework and ideally on a separate data set. Here, for each PFO the PID observables are extracted as described above, after which the weight files are used to determine the reconstructed best PID.
This information is added to the PFO and stored.
Optionally, automatic performance assessment plots can be generated, showing the relation between the true and reconstructed PID.

\section{Results of PID Combination with CPID}

In the following, two example results with CPID are shown.
The data used is single charged particles, electrons, muons, pions, kaons and protons, simulated and reconstructed with the ILD detector \cite{ILD_IDR}.
The particles' distribution is flat in $log(p)$ and isotropic.
The training is split into 12 bins in $p$ in order to ease to resolve the momentum dependence of the PID observables. A BDT is trained in each bin, using the corresponding PID observables as well as $p$ and $\theta$,  the angle with respect to the beam direction.

In order to assess PID performance, the most common quantity shown is the pion-kaon separation, generally defined as distance between pion and kaon bands in an observable normalised by the width of that band.
While this works well for Gaussian bands, e.g.\ for dE/dx, non-Gaussian distributions present a problem.
In particular BDT output scores are strongly asymmetric and using the mean and RMS of their distribution or $\mu$ and $\sigma$ of a Gaussian fit would overestimate the performance.
Instead, the so-called p-value method is applied:
When looking at the signal and background efficiency distributions, the working point is chosen so that the overlap integral is the same on both sides, i.e.\ signal loss is equal to background pick-up. This integral value is the p-value. Note that this working point is equivalent to the point where the ROC curve (signal efficiency vs.\ background rejection) intersects the $x=y$-line.
The p-value is then transferred to two Gaussians, by placing two normal distributions at such a distance that their overlap is again this value. These Gaussians now have well-defined $\mu$ and $\sigma$ and a conventional separation power $S$ is calculated.
Specifically, here the following definition is used:
$S = \frac{|{\mu_1-\mu_2}|}{\sqrt{\sigma_1^2+\sigma_2^2}}$.\\
This p-value approach is agnostic regarding the underlying PID technology and thus allows to compare different technologies including at different colliders in a coherent way.

\begin{figure}[!htp]
  \centering
  \begin{subfigure}{0.45\textwidth}
    \includegraphics[width=\textwidth,keepaspectratio=true]{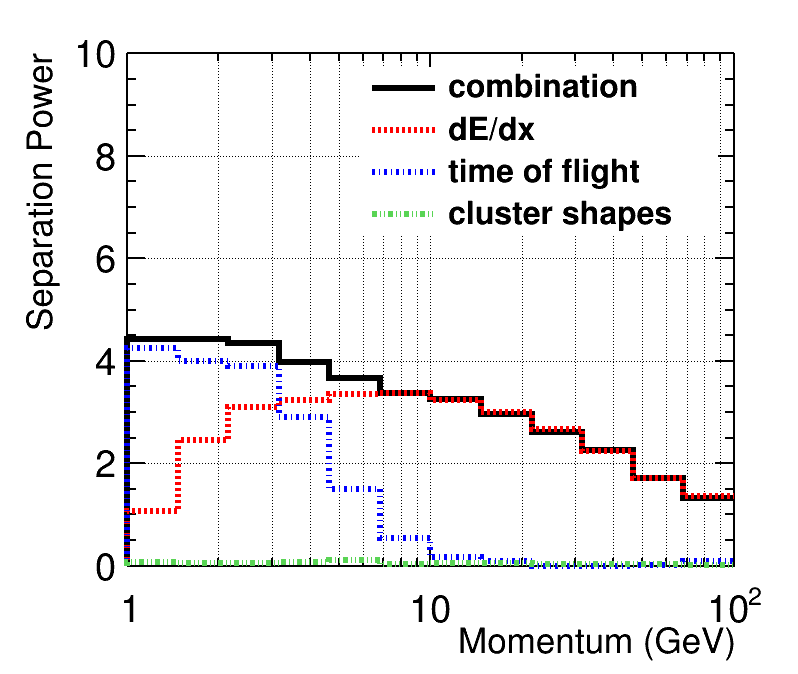}
    \caption{Pion-kaon separation power derived with CPID with different observables as well as their combination.}
  \end{subfigure}
  \hspace{.05\textwidth}
  \begin{subfigure}{0.45\textwidth}
    \includegraphics[width=\textwidth,keepaspectratio=true]{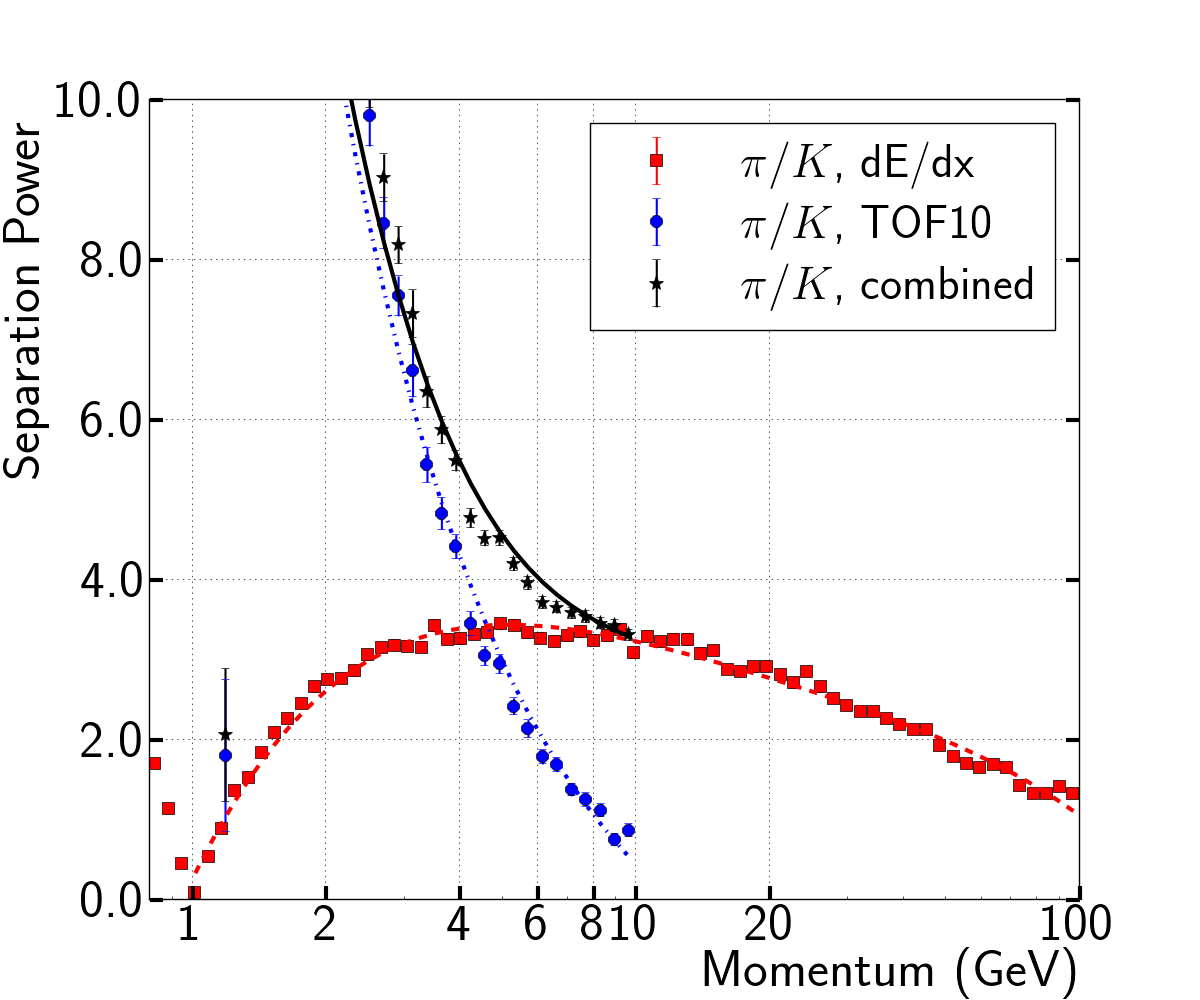}
    \caption{Pion-kaon separation power with dE/dx and TOF at ILD \cite{ILD_IDR}, extracted from \autoref{fig:PID_Perf_Exm} (a).}
  \end{subfigure}
  \caption{Pion-kaon separation with CPID and previously published by ILD.}
  \label{fig:SP_pika}
\end{figure}

The first example is the combination of different PID observables in ILD, namely dE/dx, TOF as well as an assessment of the calorimeter cluster shapes by the particle flow algorithm, which returns either electron, muon, photon or neutron.
The full CPID chain is run once for each of these three methods individually and then once with all three of them active.
The resulting pion-kaon separation power is shown in \autoref{fig:SP_pika}.
While dE/dx shows the typical curve with a maximum between 5 and 10 GeV and a zero point at 1 GeV, where the Bethe-Bloch curves of the two species cross.
TOF has a good separation at low momenta and drops quickly above 5 GeV.
As expected, PID from cluster shapes can not distinguish between hadron species.
The combination is consistent with a naive square-root-addition of the individual contributions, confirming that the observables are un-correlated.
When compared to the corresponding separation power from ILD, it is noteworthy that the TOF separation power flattens out at low momenta instead of continuing to increase, despite the data being the same. This is an effect of the p-value assessment, since for the ILD plot the TOF separation was determined by using Gaussian fits of the underlying $\beta$ distribution, which overestimates the separation power.

\begin{figure}[!htp]
  \centering  
	\begin{subfigure}{0.45\textwidth}
    \includegraphics[width=\textwidth,keepaspectratio=true]{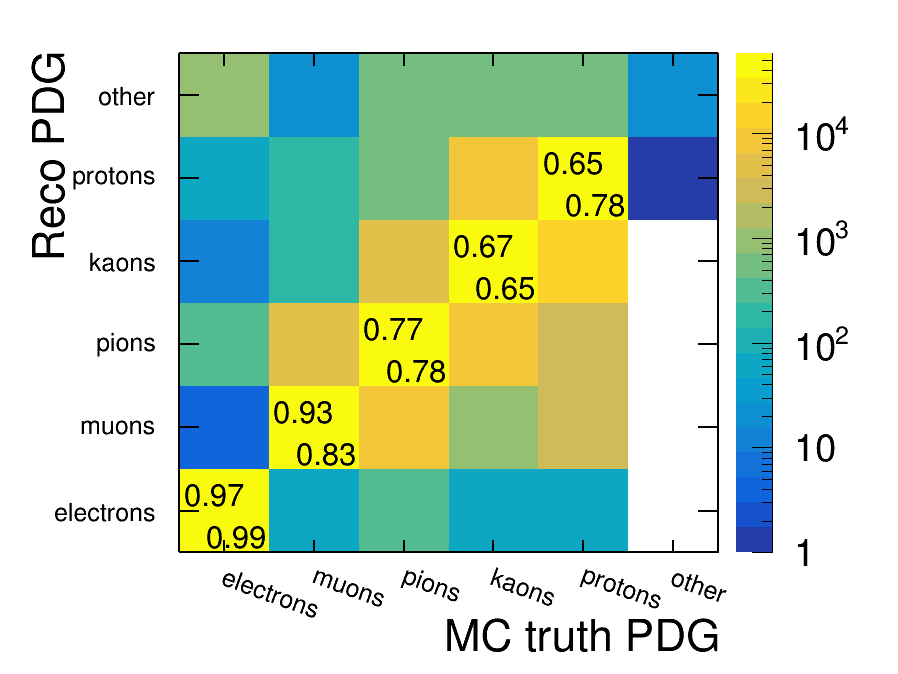}
    \caption{Confusion matrix with CPID, using dE/dx and cluster shapes.}
  \end{subfigure}
  \hspace{.05\textwidth}
  \begin{subfigure}{0.45\textwidth}
    \includegraphics[width=\textwidth,keepaspectratio=true]{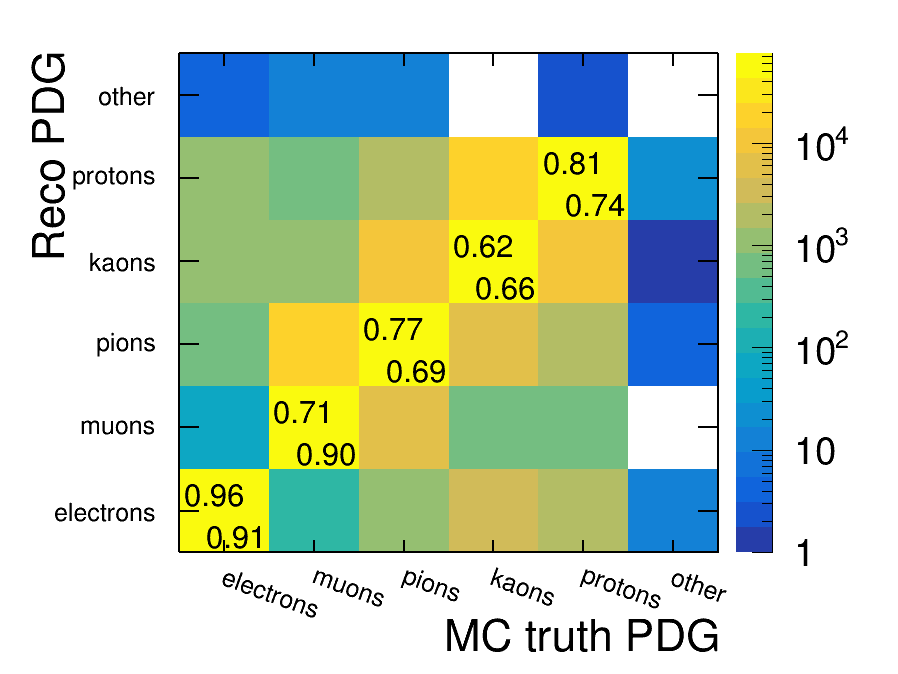}
    \caption{Confusion matrix with the LikelihoodPIDProcessor, using dE/dx and cluster shapes.}
  \end{subfigure}
  \caption{Comparison of confusion matrices of charged particle identification with the new algorithm and the one currently used in ILD.}
  \label{fig:CM_Comp}
\end{figure}

\begin{figure}[!htp]
  \centering
  \includegraphics[width=0.45\textwidth,keepaspectratio=true]{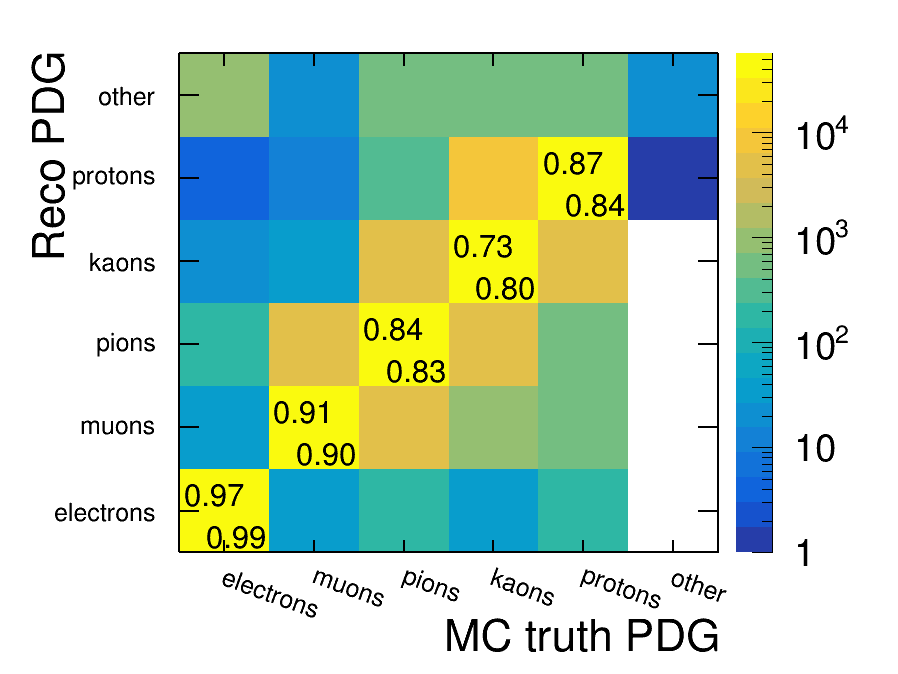}
  \caption{Confusion matrix with CPID, using dE/dx, cluster shapes and TOF.}
  \label{fig:CM_CPIDTOF}
\end{figure}

The second example uses a confusion matrix for performance assessment, see \autoref{fig:CM_Comp}. Here, for the to-be-identified 5 charged particles the true ID is compared with the reconstructed one, which is defined as the one with the highest BDT score for a particle. In addition, on each diagonal square the corresponding value for efficiency (purity) is written on the top left (bottom right).
For each species the same number of particles has been used.
The PID observables used are dE/dx and cluster shapes and CPID is compared to the algorithm currently used in the ILD reconstruction chain for PID combination, the so-called LikelihoodPIDProcessor.
The confusion matrix shows that most particles are identified correctly and the efficiency and purity values are overall comparable between the algorithms.
This indicates that the fairly basic BDT used in CPID is already competitive with the conventional method currently in use. Moreover, \autoref{fig:CM_CPIDTOF} shows the impact the addition TOF has on the confusion matrix. The improvement is clearly visible and  all efficiency and purity values now exceed the ones of LikelihoodPID.
This is not surprising, since TOF was not included in that algorithm. Instead it highlights the flexibility of CPID: While it would have been a major undertaking to integrate TOF into the LikelihoodPIDProcessor, in CPID it can be done by adding few lines in the steering file, once a TOF module has been constructed in the well defined interface.

\section{Conclusion}

Particle identification has become a topic of interest in a large number of physics studies for future colliders.
A new common framework for PID at future HTE factories, CPID, is under development and first results have been presented.
The use of the ILD detector model shows its potential: the performance is comparable to the algorithm currently in use and CPID allows to easily include additional observables.
The application to other HTE detectors is being worked on. For this, the initial implementation in iLCSoft, though usable via an existing wrapper, should be replaced by one native in key4HEP.
Finally, this framework will allow to help optimise detector designs and also to compare physics results between different colliders with a coherent performance assessment.
The p-value approach is such a coherent assessment, but it depends on the choice of one particular working point. More generic assessments are under development, in particular taking the underlying particle spectra into account.

\bibliographystyle{abbrv_mod}
\bibliography{References_reduced}
\end{document}